\documentclass[preprint,showpacs,superscriptaddress]{revtex4}

\usepackage{amssymb,amsfonts,graphicx}
\usepackage{graphicx}
\usepackage{dcolumn}
\usepackage{bm}

\begin{document}

\title{Effects of time dependency and efficiency on information flow in financial markets}

\author{Cheoljun Eom}
\affiliation{Division of Business Administration, Pusan National University, Busan 609-735, Republic of Korea}

\author{Woo-Sung Jung}
\affiliation{Center for Polymer Studies and Department of Physics, Boston University, Boston, MA 02215, USA}

\author{Sunghoon Choi}
\affiliation{Division of Business Administration, Pusan National University, Busan 609-735, Republic of Korea}

\author{Gabjin Oh}
\affiliation{Department of Physics, Pohang University of Science and Technology, Pohang 790-784, Republic of Korea}

\author{Seunghwan Kim}
\affiliation{Department of Physics, Pohang University of Science and Technology, Pohang 790-784, Republic of Korea}
\affiliation{Asia Pacific Center for Theoretical Physics, Pohang 790-784, Republic of Korea}

\date{\today}

\begin{abstract}
We investigated financial market data to determine which factors affect information flow between stocks. Two factors, the time dependency and the degree of efficiency, were considered in the analysis of Korean, the Japanese, the Taiwanese, the Canadian, and US market data. We found that the frequency of the significant information decreases as the time interval increases. However, no significant information flow was observed in the time series from which the temporal time correlation was removed. These results indicated that the information flow between stocks evidences time-dependency properties. Furthermore, we discovered that the difference in the degree of efficiency performs a crucial function in determining the direction of the significant information flow.
\end{abstract}


\maketitle

\section{\label{sec:introduction}Introduction}
Recently, researchers have become interested in the information flow occurring between financial assets or markets in an effort to understand the nature of the interaction between assets and the pricing mechanism in markets \cite{han90,milonas86,aydi98,hamao90,king90,jung06,jung08,pantron76,maslov01,granger69,engel87,eun89,shannon94,mantegna99,plerou99,schreiber00,marschinski02,kullmann02,onnela03,toth06,kumar07,kwon08}. The relationship between spots and derivatives has represented the normal course of study, particularly the manner in which the derivatives that transact with future prices affect the spots \cite{han90,milonas86}. The information flow from developed markets to emerging markets is also an issue in which active research is being conducted \cite{aydi98,hamao90,king90,jung06,jung08}. In addition, the information flow with regard to synchronization, integration and segmentation between financial markets by internal and external events has been assessed \cite{pantron76,maslov01}. Previous studies have attempted to analyze financial data using statistical method including the Granger causality test, the VAR (vector-autoregressive) model, and the GARCH (generalized autoregressive conditional heteroskedasiticity) type \cite{granger69,engel87,eun89}. However, studies regarding the factors that significantly affect information flow have proven insufficient. Therefore, we have attempted to determine empirically which factors are crucial to the information flow, considering particularly the following factors: the time-dependency property, and differences in the degree of efficiency.

According to the results of previous studies, the financial time series is time-dependent, and the time sequence exerts a significant effect on the information flow. That is to say, in financial markets, there exist many internal and external events which, as time passes, induce price changes via the interactions between stocks at the times that these events occur. In other words, the time scale of return performs a crucial function in the information flow. We have noted that the time scale of return corresponds to the time intervals, particularly when the prices are converted into the returns. Also, the efficiency of information is crucial to the pricing mechanism \cite{fama70}. The price change in a given individual stock differs from that of others, even though the same information is both instantaneously and fully reflected. This suggests that the degree of efficiency differs for each stock. Therefore, the degree of efficiency significantly affects the information flow. The efficiency we assessed in this study is based on the weak-form efficient market hypothesis (EMH), which assumes that the similarity of past price change patterns are useful in predicting future price changes. We have utilized the approximate entropy (ApEn) method in order to observe the randomness in the time series \cite{pincus91}. The ApEn method quantitatively calculates complexity, randomness, and prediction power. As the frequency of similarity patterns in the price changes is high, both the randomness and the ApEn remain low. Previous studies have argued that the ApEn evidences significant information by which the degree of efficiency can be measured \cite{pincus04,kim05,oh07}, and the predictive power and ApEn correlated negatively \cite{oh08}.

We have investigated individual stocks traded in the stock markets of Korea, Japan, Taiwan, Canada, and the USA. The entire interactions between stocks are considered, such that the number of interactions is $N(N-1)/2$, where $N$ is the number of stocks. This technique provides sufficient information flow in the market, allowing us to discover the characteristics of information flow within the context of the whole market. We detected a negative relationship between the time scale of return and the frequency of significant information flow, which supports the notion that the information flow between stocks evidences a time-dependency property. Also, we discovered that the difference in the degree of efficiency between stocks performs a crucial function in determining the direction of the information flow.

In the next section, we describe the data and the methods of the test procedures employed herein. In section \ref{sec:result}, we present the results obtained in accordance with our established research aims. Finally, we have summarized the findings and conclusions of this study.

\section{\label{sec:dataandmethods}Data and methods}
\subsection{\label{sec:data}Data}
We have assessed the daily closing prices of individual stocks traded in Korea, Japan, Taiwan, Canada, and the USA over a period of 15 years, from January 1992 to December 2006. However, we have excluded the industries which include four individual stocks or less, in order to obtain sufficient statistical analytical features. Therefore, we utilized the daily prices of 95 stocks listed on the KOSPI 200 market index of the Korean stock market, 175 stocks traded in the Nikkei 225 of the Japanese stock market, 132 stocks on the Taiwanese stock market index, 67 stocks on the TSX of the Canadian stock market, and 359 stocks in the S\&P 500 market index of the American stock market. In addition, in order to observe sufficient information flow between stocks, we considered the whole links between stocks, in accordance with the formula . The numbers of whole links between stocks for each country are as follows: 4,465 ($N=95$) links for Korea, 15,225 ($N=175$) links for Japan, 8,646 ($N=132$) links for Taiwan, 2,211 ($N=75$) links for Canada and 64,261 ($N=359$) links for the USA. The returns,$R(t)=\ln(P(t))-\ln(P(t-1))$ , are calculated using the logarithmic change of the price, $P(t)$, in which  is the stock price at $t$ day.

\subsection{\label{sec:method1}The Granger causality model in the information flow}
We investigated the empirical evidence in an effort to determine which factors affect the information flow between stocks. In order to find these factors, we have established the three following steps. First, we calculated the various returns according to the changes in the time scale; second, we calculated returns in which the various time scales apply to the causality model, considering the changes of lag length of the past data as independent variables of the model; and third, we evaluated the observed results of the information flow.

In the first step, we created a return series in accordance with the changes in the time scale, as considering the time-dependency property as a possible factor. The time scale $k$ refers to the time intervals, when the prices are converted into the returns. The time interval varies from 1 day to 5 days ($k=1,\ldots,5$). The return $R_k$ corresponds to the time scale $k$ and is calculated by$R_k=\sum_{t=1}^{k}R(t)$.

In the second step, we employed the Granger causality model to determine the direction of the significant information flow. In this model, we determined the lag length $l$ of the past data as independent variables, $X_{t-l}$ and $Y_{t-l}$, in order to explain the current price changes as the dependent variables, $X_t$ and $Y_t$. Using each return with various time scales, we have assessed the significant information flows as $X \Rightarrow Y$ and/or $Y \Rightarrow X$, considering the lag length of the past data change, $l=1,\ldots,5$, which can be defined as:

\begin{eqnarray}
\label{eq:1}
X \Rightarrow Y : &Y_t=C+\sum_{l=1}^{L} \alpha_l Y_{t-1} + \sum_{l=1}^{L} \beta_l X_{t-1}\\
&(\textrm{The null hypothesis }H_0 : \textrm{X does not Granger-cause Y})\nonumber
\end{eqnarray}
and
\begin{eqnarray}
\label{eq:2}
Y \Rightarrow X : &X_t=C^*+\sum_{l=1}^{L} \alpha^*_l X_{t-1} + \sum_{l=1}^{L} \beta^*_l X_{t-1}\\
&(\textrm{The null hypothesis }H_0 : \textrm{Y does not Granger-cause X})\nonumber
\end{eqnarray}

\noindent where $C$, $\alpha$ and $\beta$ were the estimated coefficients, the minimum and maximum of the lag length. We utilized Granger F-statistics (the Wald statistics) for each lag length with a statistical significance level of 5\% in order to determine the degree of significant information flow. Considering both the time scale and the lag length, we have statistically classified the significant information flow into three types. If both Eq. \ref{eq:1} and Eq. \ref{eq:2} are significant, the information flow results in the mutual exchange of information, $X\Leftrightarrow Y$. When either Eq. \ref{eq:1} or Eq. \ref{eq:2} is significant, we observe a one-way direction of information, $X\Rightarrow Y$ or $Y \Rightarrow X$. As both Eq. \ref{eq:1} and Eq. \ref{eq:2} are insignificant, there is no exchange of information, which can be expressed as $X\parallel Y$.

In the third step, we created measurements to evaluate the significant information flow observed in the second step since we considered both the time scale ($k=1,\ldots ,5$) and the lag length ($l=1,\ldots ,5$) in the model. Therefore, we utilized the frequency ratio (FR) of the significant information flow, which is defined as:

\begin{equation}
\textrm{FR}=\frac{\textrm{F}_\textrm{q}}{N(N-1)/2}
\end{equation}
\noindent where $\textrm{F}_\textrm{q}$ represents the frequency of significant information flow for each of the three cases; $X\Leftrightarrow Y$, $X\Rightarrow Y$ or $Y \Rightarrow X$, and $X\parallel Y$. As we investigated the changes in the information flow as functions of the time scale, we were able to confirm whether or not the information flow evidenced the time-dependency properties.

\subsection{\label{sec:method2}The approximate entropy}
The ApEn was utilized to observe the randomness in the time series so that we could quantitatively calculate complexity, randomness, and prediction power. The ApEn also assesses the degree of efficiency. The ApEn is defined as

\begin{equation}
\label{eq:4a}
\textrm{ApEn}(m,r)=\Phi^m(r)-\Phi^{m+1}(r)
\end{equation}

\noindent where $m$ represents the embedding dimension and $r$ is the tolerance of similarity.  The $\Phi^m(r)$ is given by

\begin{equation}
\label{eq:4b}
\Phi^m(r)=\frac{1}{(N-m+1)}\sum_{t=1}^{(N-m+1)} \ln \left[C_i^m(r)\right]
\end{equation}

\noindent where $C^m_i=B_i(r)/(N-m+1)$ and $B_i(r)$ is the number of data pairs within the tolerance of similarity $r$. Additionally, we calculated the similarity in the time series of each price change pattern ($u(k)$, $k=1,\ldots ,m$) by the distance $d[x(i),x(j)]$ between two vectors, $x(i)$ and $x(j)$ defined by

\begin{equation}
\label{eq:4d}
B_i=d[x(i),x(j)] \le r
\end{equation}
and
\begin{equation}
\label{eq:4e}
d[x(i),x(j)]=\max_{k=1,\ldots,m} (|u(i+k-1)-u(j+k-1)|).
\end{equation}

\noindent
In the above equations, the ApEn compares the relative magnitude between repeated pattern occurrences for the embedding dimensions $m$ and $m+1$. The ApEn becomes smaller as the frequency of similar price change patterns for the embedding dimension $m$ becomes equal to those for the embedding dimension $m+1$. When both frequencies are equal, the ApEn is 0. Therefore, the ApEn is small when the frequency of the similar price change pattern is high. The time series data evidences a low degree of randomness, and the efficiency of the financial time series becomes low. In this study, we have utilized the embedding dimension $m=2$ and the tolerance of similarity $r=20\%$ of the standard deviation of the time series.

\section{\label{sec:result}Results}
\subsection{\label{sec:result1}The time-dependency property}
\begin{figure}
\includegraphics[width=1.0\textwidth]{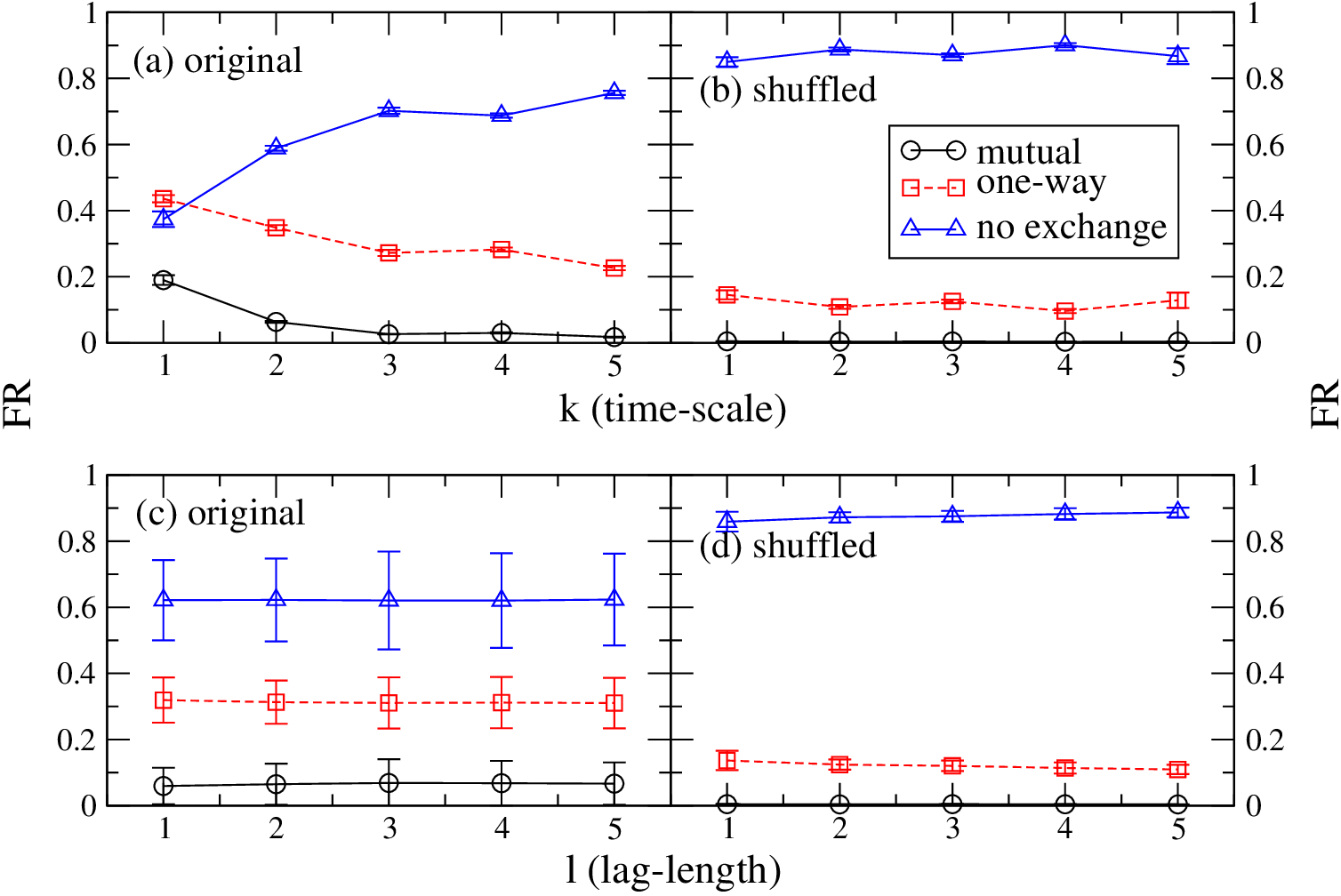}
\caption{\label{fig:1} (Color online). The frequency ratio (FR) as functions of the time scale $k$ and the lag length $l$ for the Japanese market. Both $k$ and $l$ range between 1 and 5. The original data was analyzed in (a) and (c), and the shuffled data in (b) and (d). The circle corresponds to the mutual exchange, the square to the one-way direction exchange, and the triangle to no exchange, respectively.}
\end{figure}

We attempted to determine whether the information flow changes with the time scale. Fig. \ref{fig:1} shows the FR for 175 stocks listed on the Nikkei 225 of the Japanese stock market. The lag length $l$ and the time scale $k$ were shown to vary from 1 day to 5 days. The shuffled data was analyzed in order to eliminate the temporal correlation in the time series. The shuffled data evidences the same statistical properties, including the mean, variance, skewness, and kurtosis with the original time series. In Figs. \ref{fig:1}(a) and (b), the X-axis denotes the time scale, whereas the Y-axis refers to the  of the information flow. The X-axes of Figs. 1(c) and (d) correspond to the lag length. The circles, squares, and triangles indicate the mutual exchanges, the one-way direction, and no exchanges of the information, respectively. Figs. \ref{fig:1}(a) and (c) represent the results from the original data, and Figs. \ref{fig:1}(b) and (d) show those for the shuffled data.

\begin{figure}
\includegraphics[width=1.0\textwidth]{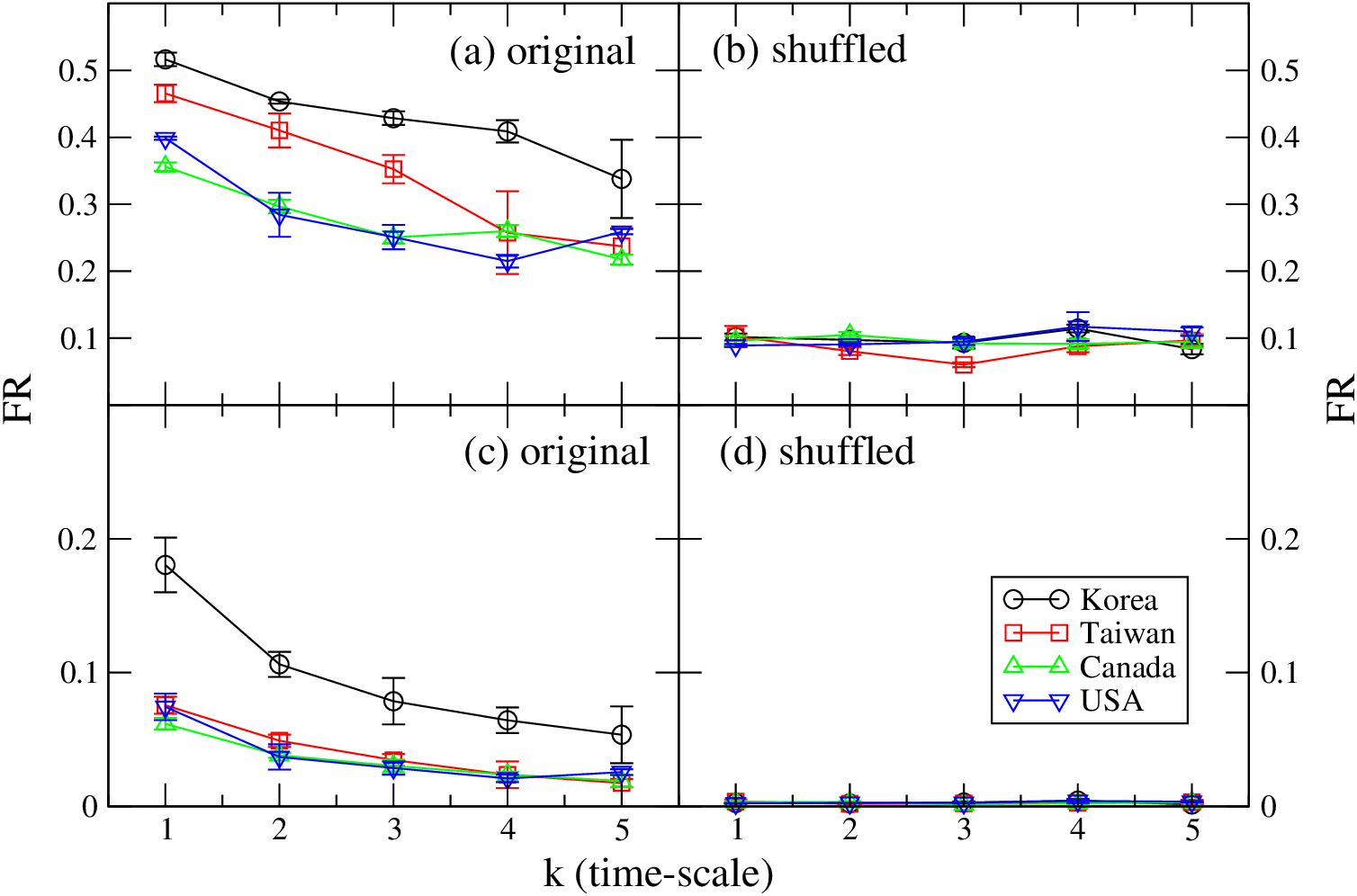}
\caption{\label{fig:2} (Color online). The frequency ratio (FR) as functions of the time scale $k$  for the Korean (circle), Taiwanese (square), Canadian (triangle up), and US markets (triangle down). The original data was assessed in (a) and (c), and the shuffled data in (b) and (d). (a) and (b) represent the one-way direction exchange, and (c) and (d) the mutual exchange.}
\end{figure}

In Fig. \ref{fig:1}, we determined that the information flow observed by the Granger causality model is influenced significantly by the time scale. The FRs for the mutual exchange of information and the one-way direction of information decrease as the time scale $k$ increases from 1 day to 5 days, whereas that for no exchange of information increases. However, we have not discovered any significant trend as the lag length $l$ increases. This means that the information flow evidences a time-dependency property. Fig. \ref{fig:2} represents the  as a function of  for other markets, including Korea, Taiwan, Canada, and the USA. Figs. \ref{fig:2}(a) and (b) show the one-way direction of information, and (c) and (d) represent the mutual exchange of information.

Figs.  \ref{fig:1}(b), (d) and  \ref{fig:2}(b), (d) correspond to the FRs for the shuffled data with no time correlations. In those figures, we determined that the information flow is quite sensitive to the time sequence. The FRs for the mutual exchange and the one-way direction of information is very small, regardless of the values of  $k$ and $l$. However, we noted that the FR for no exchange of information was higher in Figs. \ref{fig:1}(b) and (d). Therefore, we were unable to detect any significant information flow in cases in which the time sequence was removed.

\subsection{\label{sec:result2}The difference in the degree of efficiency}
Next, we attempted to determine empirically whether information flow between stocks was affected by differences in the degree of efficiency between stocks. The price reaction of individual stocks differed in accordance with whether the new information was instantaneously and fully reflected in the price. The fact is, the degree of efficiency differs among each stocks. Therefore, the difference in the degree of efficiency between stocks may influence the interactions between stocks, and may exert a significant effect on information flow. We have tested the assumption that the more efficient stocks reacts quickly against the information, such that the price also changes more quickly than does the less efficient stock. Also, the more efficient stock with relatively more sufficient information delivers the information to the less efficient stock. Finally, the information flow is correlated with the difference in the degree of efficiency.

In order to quantify the degree of efficiency, we utilized the ApEn method, which quantifies the degree of complexity, irregularity, and randomness in the time series. The ApEn is calculated on the basis of the degree of similarity patterns in the time series. Therefore, the ApEn is related to the EMH, and can thus be used to quantify the degree of  efficiency in the information flow. If the time series evidences a high degree of randomness and complexity, the ApEn is large, whereas the ApEn of the random time series is quite high.

\begin{figure}
\includegraphics[width=1.0\textwidth]{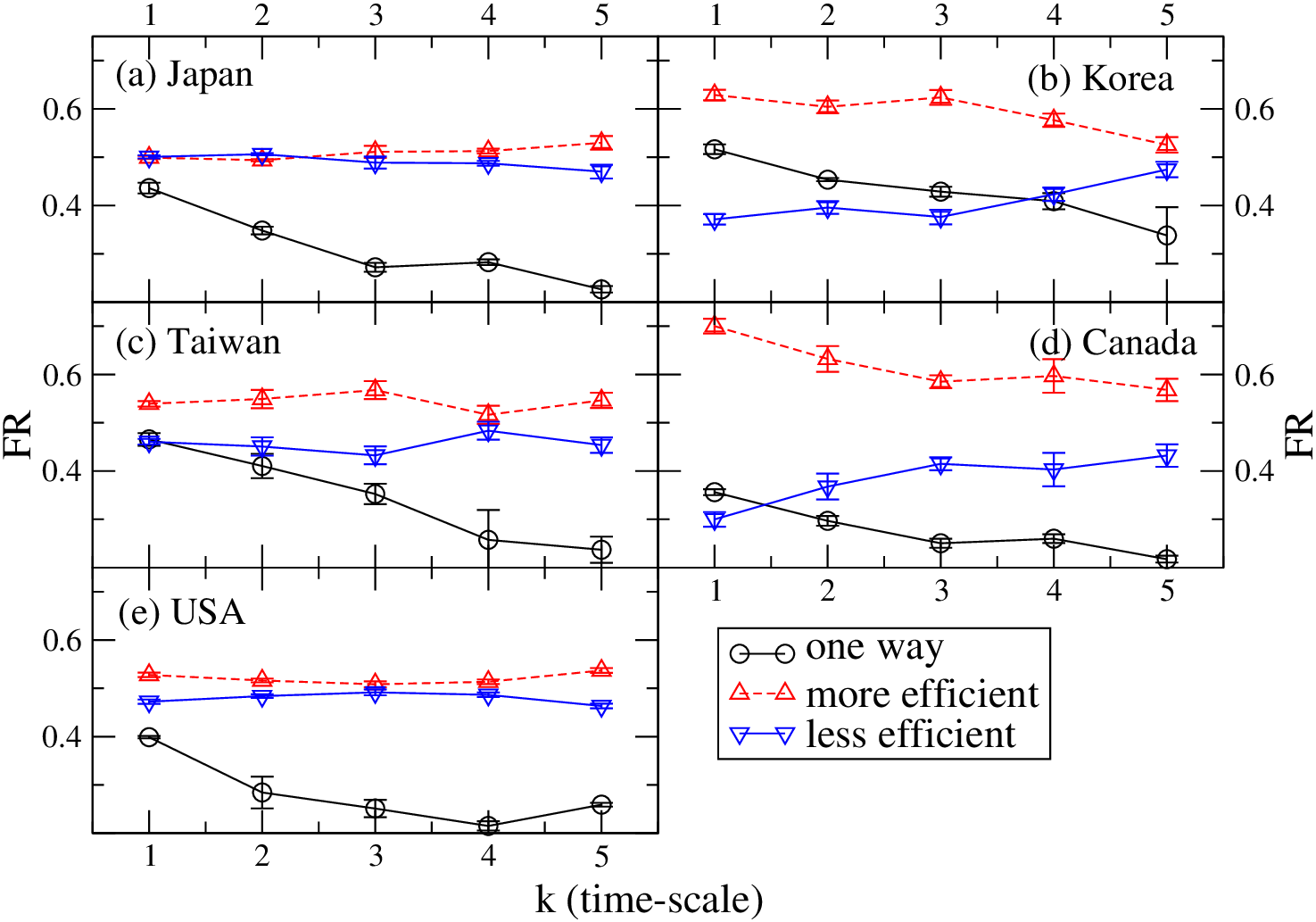}
\caption{\label{fig:3} (Color online). The frequency ratio (FR) for $X^E \Rightarrow Y^E$ (triangle up) and $Y^E \Rightarrow X^E$ (triangle down) efficient stocks when $X^E$ is the more efficient stock in comparison with $Y^E$. The one-way direction case (circle) is provided from Figs. \ref{fig:1} and \ref{fig:2} for purposes of comparison.}
\end{figure}

Fig. \ref{fig:3} demonstrates the frequency ratios for $X^E \Rightarrow Y^E$ (triangle up) and $Y^E \Rightarrow X^E$ (triangle down) efficient stocks when $X^E$ is the more efficient stock as compared with $Y^E$. The one-way direction of information cases (circle) from Figs. \ref{fig:1} and \ref{fig:2} are provided for purposes of comparison. In Fig. \ref{fig:3}, the FR for $X^E\Rightarrow Y^E$ was shown to be higher than that for the opposite direction $Y^E\Rightarrow X^E$ for all of the analyzed markets. This means that the more efficient stock $X^E$ exerts the driving force of the information flow into the less efficient stock $Y^E$. In particular, the difference between $X^E\Rightarrow Y^E$ and $Y^E\Rightarrow X^E$ for the Korean and Canadian markets are relatively obvious and clear. However, the reason why those markets' features are more clear remains an open question, and will be the subject of our future work. On the basis of these results, we discovered that the difference in the degree of relative efficiency between stocks influences the direction of the information flow, and thus plays an important factor.

\section{\label{sec:conclusions}Conclusions}
We have investigated stock market data in order to determine which factors perform crucial functions in the information flow between stocks. The time dependency property and the difference in the degree of efficiency were assessed using the Granger causality model and the approximate entropy.

We found that the information flow observed via the Granger causality model was influenced significantly by the time scale. The time scale and the frequency of the significant information flow were correlated negatively. However, we were not able to detect a significant information flow when the temporal time correlation was removed from the time series. We also discovered that the difference in the degree of efficiency played a crucial role in determining the direction of the information flow. We confirmed that the information flowed from the more efficient stock to the less efficient stock. These findings indicated that the information flow evidences a time-dependency property, and is influenced by the difference in the degree of efficiency. In our opinion, when researchers further assess the information flow between assets or markets, they should consider the time-dependency property and the difference in the degree of efficiency.


\newpage
\clearpage

\end{document}